\documentclass[showpacs,prb,letterpaper,floatfix,nobalancelastpage,twocolumn]{revtex4}
\usepackage{dcolumn}
\usepackage{bm}
\usepackage{amsmath}
\usepackage{bm,braket}
\usepackage{color}
\usepackage{float}
\usepackage{graphicx,subfigure}

\begin{document}


\title{Nonlinear photon transport in a semiconductor   waveguide-cavity system containing a single quantum dot: Anharmonic cavity-QED regime}
\author{S. Hughes}
\email{shughes@physics.queensu.ca}
\author{C. Roy}
\affiliation{Department of Physics, 
Queen's University,
 Kingston, Ontario,  Canada K7L 3N6}
\begin{abstract}
We present a semiconductor master equation technique to study the input/output characteristics
of  coherent photon transport in a semiconductor     waveguide-cavity system containing a  single  
quantum dot.
We use this approach to investigate the effects of   photon  propagation  and anharmonic cavity-QED for various dot-cavity interaction strengths, including weakly-coupled, intermediately-coupled, and strongly-coupled regimes. We demonstrate that  for mean photon numbers  much less than 0.1,  the commonly adopted weak excitation (single quantum) approximation  breaks down, even in the weak coupling regime. As a measure of the anharmonic multiphoton-correlations, we compute the Fano factor and the   correlation error associated with making a semiclassical approximation. We also explore the role of electron--acoustic-phonon scattering  and find that phonon-mediated scattering   plays a qualitatively important role on the light propagation characteristics.
As an application of the theory,  we  simulate a  conditional  phase gate
at a phonon bath temperature of $20~$K in the strong coupling regime.
\end{abstract}

\maketitle

\section{Introduction}
\label{intro}
The ability to couple waveguides and cavities offers exciting opportunities
for integrated quantum optical devices using solids~\cite{Cirac:PRL97,Vahala:Nature03,Wang:PRL05}.
In particular, planar photonic crystals offer a technology platform, where quantum bits (qubits) can be
manipulated from quantum dots (QDs) placed at field antinode positions within the
cavity or waveguide~\cite{England:OE07,OurReview:OE,ObrianReview:Nature10}.
Integrated semiconductor micropillar systems also show great promise
for quantum optical applications~\cite{pillar1,pillar2}, working at the
few photon level.

Recently, there have been several successful demonstrations
of coherent light {propagation} effects in various semiconductor systems,
including planar photonic crystals
and micropillars. Bose {\em et al.}~\cite{Bose:OE10}
 measured the exciton-induced doublet (polariton splitting)
through waveguide mode transmission in a photonic crystal waveguide-cavity system~\cite{hughes:2004} [cf.~Fig.~1(a)], while
Loo {\em et al.}~\cite{Loo:APL10} probed the strong coupling in a micropillar via coherent reflection [cf.~Fig.~1(b)];
Young {\em et al.}~\cite{Young:arXiv10} demonstrated first steps toward
a conditional phase gate  using light
reflection from a micropillar.
Common to the analysis of all of these experiments has been
the application of the {\em weak excitation approximation} (WEA), where at most only one quantum is assumed.
 For example,
Ref.~\onlinecite{Young:arXiv10}
suggested that their experiments were likely at the ``single photon level'' for less than 0.1 photons 
per cavity lifetime, so they applied
a WEA solution. The same assumptions are tacitly made by many other
groups, in excitation regimes where the mean photon number is 
well below  those
associated with saturating the QD exciton~\cite{Englund:Nature07}.
These useful formalisms  have been very successful and certainly help to clarify the
basic physics of low-intensity photon transport.

For increasing field strengths, however, the validity for the WEA becomes questionable, and there can be quantum
nonlinearities in the system due to multiphoton correlations.
Giant optical nonlinearities were  studied by Auff\'eves-Garnier {\em et al.}~\cite{Auffeves:PRA07}; their semiclassical approach adiabatically eliminated the cavity mode and included effects outside the  WEA (the ``Purcell regime''); naturally with such a semiclassical approach, there is no influence from the higher lying levels of the {\em anharmonic} 
Jaynes-Cumming (JC) ladder, so it cannot be applied in the strong coupling regime.
Most photon transport  approaches also neglect the details of electron-acoustic phonon scattering 
\cite{besombes,Frank:PRB08,HohenesterPRB:2010,hughes1,Calic:PRL11}---apart
from the
inclusion of a Lorentzian decay rate for the exciton, i.e., broadening of the the 
zero phonon line (ZPL).
Recently, several works have shown that coherent excitation of semiconductor-QD systems can easily go into the anharmonic cavity-QED (quantum electrodynamics) regime~\cite{roy_hughes,me_PRL01}.
Thus the questions arise: ($i$) To what extent can one safely employ the WEA 
for these emerging semiconductor
 waveguide-cavity and micropillar
systems? ($ii$) When are multiphoton correlation effects important (quantum nonlinear regime)? ($iii$) What is the role of electron-acoustic phonon scattering
and how does this mechanism differ from a simple-Lorentzian pure dephasing model?

In this work we  attempt to answer  these  questions for the
regime of coherent photon transport,
and show that generally one must include
multiphoton effects and phonon scattering within the theoretical formalism.
Even for weakly-coupled QD-cavity systems, higher-order quantum correlations
effects are shown to be significant. Our paper is organized as follows: In Sect.~\ref{theory}, we introduce the general theoretical technique to simulate coherent photon transport
outside both the WEA and  the semiclassical approximation.
The theory is based on a quantum master equation (ME) formalism
where cavity and dot decays are introduced by Lindblad superoperators, and phonon interactions are included from an effective phonon ME~\cite{PRX} that includes electron-phonon interactions at a microscopic level---derived using a polaron transform~\cite{PRX,roy_hughes}.
In Sect.~\ref{results},
we first present calculations with no acoustic phonon coupling, for several different QD-cavity couplings;  both the weak coupling
regime and the intermediate-to-strong coupling regime are investigated. We demonstrate that 
substantial deviations from the WEA can result, 
even for
very small mean photon numbers $\ll 0.1$. We then  modify the ME approach to
include the mechanism of electron--acoustic-phonon scattering, and study the impact of electron-phonon interactions on incoherent scattering
and on coherent renormalization of the exciton-cavity coupling rate; qualitative differences from a simple Lorentzian decay model are found. As an application of the theory, we study the transmission of light in the strong coupling regime
 and  simulate
a conditional (exciton-induced) phase gate. We conclude in Sect.~\ref{conclusions}.

\section{Theory}
\label{theory}

\begin{figure}[t!]
\centering
\includegraphics[width=0.95\columnwidth]{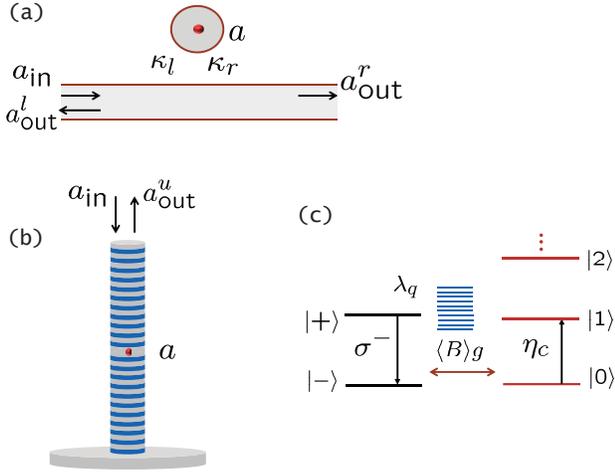}
\vspace{-0.0cm}
\caption{(Color online) Schematic of two semiconductor cavity-QED
systems containing a quantum dot: (a) waveguide-cavity system
(e.g., made up of a
planar photonic crystal) where we assume $\kappa_l=\kappa_r=2\,\kappa_0$
(where $\kappa_0$ is the  out of plane loss),  and
(b) a micropillar cavity.
(c) Dot and cavity energy levels, where the
exciton system
($\ket{+/-}$)
 interacts with an acoustic phonon bath (see text).
}
\label{fig:schematic}
\end{figure}

We wish to describe light propagation for a QD-cavity geometry,
where the input and output fields can be identified separately from the
cavity region in which the QD is assumed to be embedded.
An example waveguide-cavity system
is shown schematically in Fig.~1(a).
For a continuous wave (cw) waveguide mode of a photonic crystal system,
the classical WEA reflectivity was
previously derived  by Hughes and Kamada~\cite{hughes:2004,note1}:
\begin{align}
r_{\rm pc}(\omega) = \frac{i\omega\Gamma_c}{\omega_c^2-\omega^2-i\omega(\Gamma_c+\Gamma_0) -\omega\Sigma(\omega)},
\end{align}
where the self-energy
$\omega\Sigma(\omega)=\omega g^2/(\omega_x-\omega^2-i\omega\Gamma_x^{t})$, $\Gamma_0 \equiv 2\kappa_0$ is the cavity decay rate through vertical scattering ({\em unloaded}
cavity broadening),  $\Gamma_{c}\equiv 2 \kappa_{c}=2(\kappa_l+\kappa_r$) 
is the cavity-waveguide coupling rate (which is inversely proportional to the
group velocity of the waveguide mode~\cite{hughes:2004}),
$\omega_c$ is the cavity mode resonance,
$\omega_x$ is the target exciton resonance of the QD,
and $\Gamma_x^{t}=\gamma+\gamma'$ is the total decay rate of the exciton---including radiative 
($\gamma$) and non-radiative, pure dephasing ($\gamma'$) processes.
The total cavity decay rate is $\Gamma_c^t=\Gamma_0+\Gamma_c$ and
the exciton-cavity coupling rate, $g\propto d^2/V_{\rm eff}$, where
$d$ is the dipole moment of the exciton and
$V_{\rm eff}$ is the effective mode volume.
The corresponding transmissivity is simply $t=1+r$, and 
one can also define the reflection and transmission, through $R=|r|^2$ and $T=|t|^2$.
Similar expressions have
been derived by other groups, e.g.,
Refs.~\onlinecite{Waks:PRL06,Auffeves:PRA07,Fan:PRA09}.
With the dot on resonance with the cavity, then
a polariton doublet
coincides with the vacuum Rabi splitting which
can be observed in transmission or reflection;
 this {\em normal mode} doublet can occur even if the dot is {\em not} in the strong coupling
regime~\cite{Waks:PRL06,Auffeves:PRA07,Rakher:PRL09}, though ultimately the doublet
feature is lost at high temperatures due to phonon bath coupling~\cite{Frank:PRB08}.
The analysis of coherent reflection from a micropillar system is similar; one has
$r_{\rm \mu pill}=1-\sqrt{\eta}\, r_{\rm pc}$,
where $\eta$ is  a measure of in/out coupling
efficiency~\cite{Loo:APL10}, and one also makes the following replacements:
$\Gamma_c\rightarrow \Gamma_0$  (vertical scattering)
and
$\Gamma_0 \rightarrow \Gamma_s$ (sidewall scatter).

The above WEA formalism  does not distinguish between
radiative and pure dephasing processes of the QD exciton (which a semiclassical
approach can); nor does it take into
consideration the acoustic phonon bath  or multiphoton effects.
Ignoring the complexities of phonon scattering for now~\cite{Frank:PRB08,hughes1,Calic:PRL11}---which we will introduce  below---these analytical formulas are expected to work for only weak excitation conditions (strictly linear),
and for pure dephasing rates~\cite{Bose:OE10}, 
$\gamma' \ll g^2\Gamma_c^t/[4(\omega_c-\omega)^2+(\Gamma_c^t/2)^2]$.

To go beyond the WEA, and the semiclassical approches, 
we will use a ME approach
 where  
 exciton-photon interactions are easily included to all orders.
Referring to Fig.~\ref{fig:schematic}(a),
we relate the left/right output operators
 to the cavity mode operator through~\cite{carmichael:2005}
\begin{align}
\braket{a_{\rm out}^r(t)} & = - \braket{a_{\rm in}(t)} + \sqrt{2\kappa_c}\, \braket{a(t)}, \\
\braket{a_{\rm out}^l(t)} & = \sqrt{2\kappa_c} \, \braket{a(t)},
\end{align}
where, for a  coherent cw input state,
$\braket{a_{\rm in}}=i\eta_c/(2\sqrt{2\kappa_c})$, with 
$\eta_c$ the cavity pump rate.
Following the solution of the   ME (discussed below),
the {\em steady-state} transmissivity and reflectivity
are obtained:
$t \equiv |t|e^{i\phi_t} =  {\braket{a_{\rm out}^r}_{\rm ss}}/{\braket{a_{\rm in}}}$
and
$r \equiv |r|e^{i\phi_r} = {\braket{a_{\rm out}^l}_{\rm ss}}/{\braket{a_{\rm in}}}$ ,
where $\phi_t$ and $\phi_r$ are the phases.
Working in a frame rotating with respect to the laser pump frequency, $\omega_L$,
 the model Hamiltonian can be written as
\begin{align}
\label{sec1eq1}
H&=\hbar\Delta_{xL}{\sigma}^{+}{\sigma}^{-}+\hbar\Delta_{cL}{a}^{\dagger}{a} +
\hbar g({\sigma}^{+}{a}+{a}^{\dagger}{\sigma}^{-}) \nonumber \\
&+H^{c}_{\rm{drive}} +{\sigma}^{+}{\sigma}^{-}\sum_{q}\hbar\lambda_{q}({b}_{q}
+{b}_{q}^{\dagger})+\sum_{q}\hbar\omega_{q}{b}_{q}^{\dagger}{b}_{q}\, ,
\end{align}
where  ${b}_{q}({b}_{q}^{\dagger})$ are the annihilation and creation operators of the phonons,
 $a$ is the 
 cavity mode annihilation operator,  ${\sigma}^+,{\sigma}^-$
 are  Pauli operators of
the electron-hole pair (exciton), $\Delta_{\alpha L}\equiv \omega_\alpha-\omega_L$ ($\alpha =x,c$)
are the detunings of the exciton ($\omega_{x}$) and cavity ($\omega_{c}$) from $\omega_{L}$,
 and
$H^{c}_{\rm{drive}} = \hbar \eta_{c}({a}+{a}^{\dagger})$
 describes the
 {\em coherent} cavity drive (excited through the waveguide channel).

 To obtain the ME,
 we first transform Eq.~(\ref{sec1eq1}) to a polaron frame,
 which
formally   recovers the  independent boson model (IBM)~\cite{mahan,imamoglu,krum};
the IBM is known to  accurately describe the characteristic lineshape of
a single exciton coupled to a bath of acoustic phonons~\cite{besombes}.
Specifically, the polaron ME~\cite{imamoglu,nazir2,roy_hughes}
introduces coherent electron-phonon coupling exactly,
while incoherent phonon interactions are
treated at the level of a second-order Born approximation.
Further details og the model are discussed in Refs.~\onlinecite{nazir2,roy_hughes}.
The time-convolutionless ME takes the form~\cite{nazir2,roy_hughes}:
\begin{align}
\label{sec3eq3}
\frac{\partial \rho}{\partial t}&=\frac{1}{i\hbar}[H_{\rm sys}^{\prime},\rho(t)]+{\cal L}(\rho)
+ {\cal L}_{\rm ph}(\rho) ,
\end{align}
where the polaron-transformed system Hamiltonian is
$
H^{\prime}_{\rm sys}   =    \hbar(\Delta_{xL}-\Delta_{P}){\sigma}^{+}{\sigma}^{-}
+\hbar\Delta_{cL} {a}^{\dagger}{a}+\langle B\rangle {X}_{g}+H_{\rm drive}^c$,
with
$\langle B\rangle=\exp [ -\frac{1}{2}\int^{\infty}_{0}d\omega{J(\omega)}/{\omega^{2}}\coth(\beta\hbar\omega/2) ] (\beta=1/k_bT)$,
 $X_g = \hbar g( {a}^{\dagger}{\sigma}^{-}+{\sigma}^{+}{a})$,
and $\Delta_P=\int^{\infty}_{0}d\omega{J(\omega)}/{\omega}$. In what follows,
we will absorb the polaron shift ($\Delta_P$) into the definition of $\omega_x$.
The phonon spectral function~\cite{ota,nazir2,roy_hughes},
 $J(\omega)=\alpha_{p}\,\omega^{3}\exp (-{\omega^{2}}/{2\omega_{b}^{2}} )$,
describes electron-acoustic phonon interaction via a deformation potential coupling.

Using a Markov approximation, the incoherent phonon scattering term is  defined as~\cite{roy_hughes}
\begin{align}
\label{eq:Lph}
{\cal L}_{\rm ph}(\rho)& =
-\frac{1}{\hbar^{2}}\int^{\infty}_{0}d\tau\sum_{m=g,u}
 \left ( G_{m}(\tau) \phantom{{X}_{m},e^{-iH_{sys}^{\prime}}} \right . \nonumber \\
&\!\!\!\!\!\!\!\!\!\! \times \left . \left [{X}_{m},e^{-iH_{\rm sys}^{\prime}\tau/\hbar}{X}_{m}e^{iH_{\rm sys}^{\prime}\tau/\hbar}\rho(t)\right]
+{\rm H.c.} \right ),
\end{align}
where $X_u= -i\hbar g(a^\dagger{\sigma}^{-}-{\sigma}^{+}a)$, and $G_{g/u}(t)$ are the
polaron Green functions~\cite{mahan,imamoglu}:
$G_{g}(t)=\langle B\rangle^{2}\left (\cosh[\phi(t)]-1 \right ),
G_{u}(t)=\langle B\rangle^{2}\sinh[\phi(t)]$,
with
$\phi(t)=\int^{\infty}_{0}d\omega\frac{J(\omega)}{\omega^{2}}
\left [\coth(\beta\hbar\omega/2)\cos(\omega t)-i\sin(\omega t)\right ]$.
%
%
%
In Ref.~\onlinecite{PRX}, an effective Lindblad ME has been shown to yield
very good agreement with the full polaron ME solution above. In this
way, one defines the phonon-mediated incoherent scattering processes through
\begin{align}
\label{sec3eqfinal3}
{\cal L}_{\rm ph}(\rho)=\frac{\Gamma_{\rm ph}^{\sigma^{+}a}}{2}{\cal L}({\sigma}^{+}{a})
+\frac{\Gamma_{\rm ph}^{a^{\dagger}\sigma^{-}}}{2}{\cal L}({a}^{\dagger}{\sigma}^{-}),
\end{align}
where
${\cal L}({D})=2{D}\rho{D^{\dagger}}-{D^{\dagger}}{D}\rho-\rho{D^{\dagger}}{D}$, and the scattering rates are obtained analytically, from~\cite{PRX}
\begin{align}
\label{eq:phononrates}
\Gamma_{\rm ph}^{\sigma^{+}a/a^{\dagger}\sigma^{-}}& =2\braket{B}^2 \! g^{2}\,{\rm Re} \left [\int_{0}^{\infty}d\tau\,
e^{\pm i\Delta_{cx} \tau}\! \left (e^{\phi(\tau)}-1 \right )\right],
\end{align}
where $\Delta_{cx}=\omega_c-\omega_x$ is the cavity-exciton detuning. 
The rate $\Gamma_{\rm ph}^{a^\dagger\sigma^-}$ describes the process of cavity excitation 
and the emission of a exciton, via phonon-induced scattering,
and $\Gamma_{\rm ph}^{\sigma^+a}$ describes exciton excitation via the emission of a cavity photon.
For completeness, one can also add in the Stark shifts~\cite{roy_hughes}, but these
are found to be negligible for the regimes considered in this work.
An alternative weak--phonon-coupling theory  is presented in Ref.~\onlinecite{jelena_arka}.
Since this Lindblad--ME-form is considerably easier to work with and helps to identify the physics of the scattering processes in a more transparent way, we will use this latter form for our phonon  calculations in this paper. 

Phenomenologically,
we also
include Liouvillian superoperators~\cite{ota,roy_hughes}:
\begin{align}
{\cal L}(\rho) &=\frac{\tilde{\gamma}}{2}(2{\sigma}^{-}\rho{\sigma}^{+}
-{\sigma}^{+}{\sigma}^{-}\rho-\rho{\sigma}^{+}{\sigma}^{-}) \nonumber \\
&+\kappa_t(2{a}\rho{a}^{\dagger}-{a}^{\dagger}{a}\rho-\rho{a}^{\dagger}{a})
\nonumber \\
&+\frac{\gamma^{\prime}}{2}(
{\sigma}_{11}\rho{\sigma}_{11}-
 {\sigma}_{11}{\sigma}_{11}\rho
 - \rho\sigma_{11}{\sigma}_{11}
 ),
 \end{align}
where
$2\kappa_t=2\kappa_0+2\kappa_w\equiv \Gamma_c^t$,
$\tilde{\gamma}=\gamma\langle B\rangle^{2}$
is the radiative decay rate of the exciton, $\gamma'$ is the pure dephasing rate of the exciton, and
${\sigma}_{11}={\sigma}^{+}{{\sigma}^{-}}$.
Figure 1(c) shows a schematic of the model, which we solve
in a basis of $\ket{n}$ photons and the two QD states $\ket{-/+}$.
The WEA corresponds to
a three state model, i.e.,
only including the ground ground state and
and the first two ladder states
of the JC model.

For suitably large input fields, many JC ladder states may be involved. To  better highlight the role of quantum statistics in the system
and to help quantify the possible failure of a semiclassical
approximation, we introduce a function for computing the relative {\em correlation error}
in the expectation $\braket{a^\dagger\sigma^-}_{\rm ss}$  compared to $\braket{a^\dagger}_{\rm ss}\braket{\sigma^-}_{\rm ss}$, 
\begin{equation}
{\rm CE}=\frac{|\braket{a^\dagger\sigma^-}_{\rm ss}-\braket{a^\dagger}_{\rm ss}\braket{\sigma^-}_{\rm ss}|}
{|\braket{a^\dagger\sigma^-}_{\rm ss}|} ,
\end{equation}
where $100\, {\rm CE}$ is the percentage relative error; in a semicassical picture, of course ${\rm CE}=0$. We also
compute the mean photon variance, defined through the 
  Fano factor:
\begin{equation}
F = \frac{\braket{(a^\dagger a)^2}_{\rm ss} - \braket{a^\dagger a}_{\rm ss}^2}
{\braket{a^\dagger a}_{\rm ss}}.
\end{equation}

\section{Results}
\label{results}
\subsection{Photon transport with no acoustic phonon coupling}

\begin{figure}[t!]
\centering
\includegraphics[width=0.95\columnwidth]{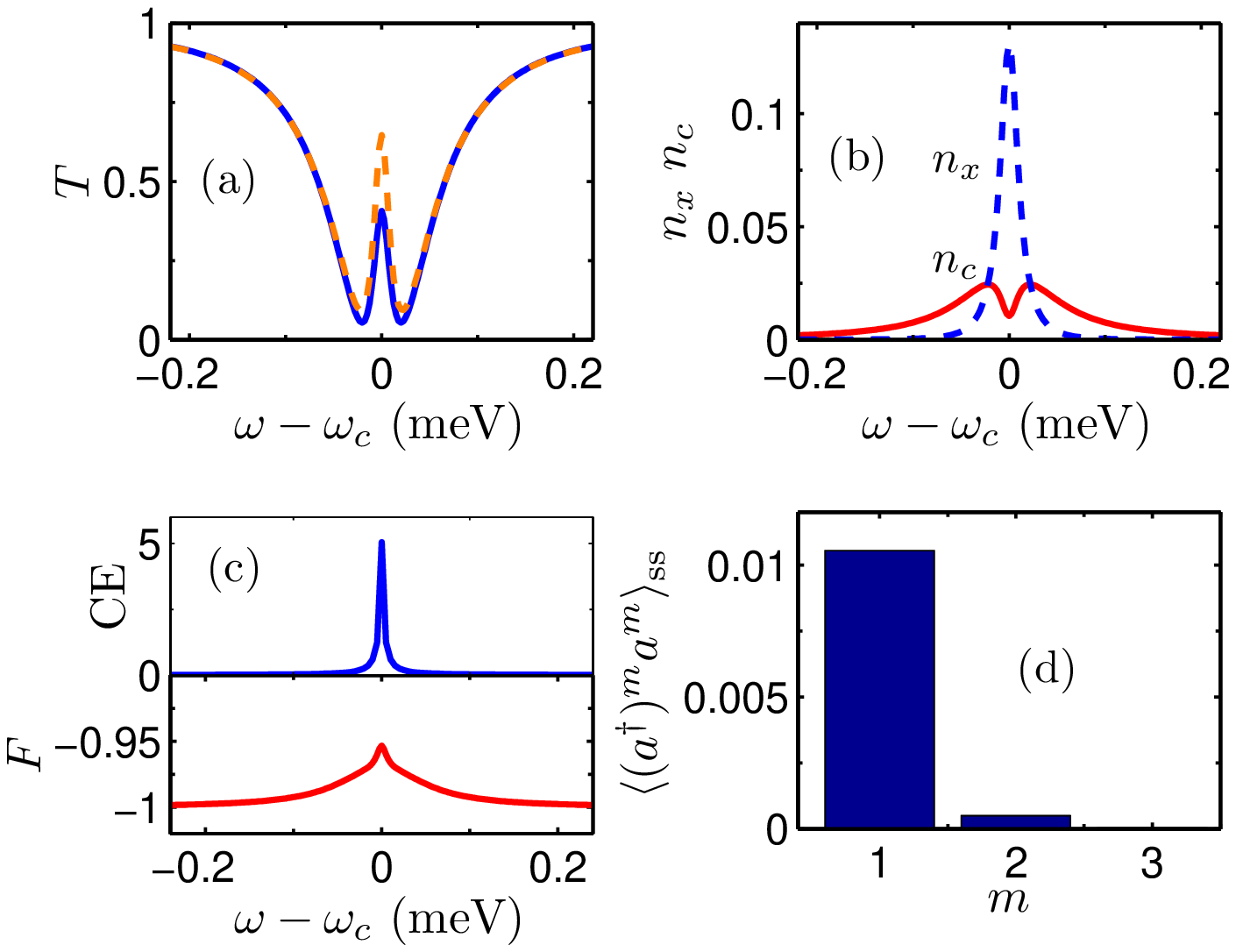}
\vspace{-0.15cm}
\caption{(Color online)
Transmission characteristics for a weakly-coupled QD--cavity-waveguide 
system,  with $g=20~\mu$eV $\approx 0.32\,\kappa_t$ and $\eta_c=0.5\,g$.
(a) Transmission with (dashed) and without (solid) the WEA approximation.
All other calculations in (b-d) use the full multiphoton approach.
(b) Steady-state  cavity (solid) and exciton (cavity) populations.
(c) Relative correlation  error of $\braket{a^\dagger\sigma^-}_{\rm ss}$ if a semiclassical
approximation was used, CE, and the Fano factor (photon number variance).
(d) Cavity photon moments
$\braket{(a^\dagger)^ma^m}_{\rm ss}$ versus $m$.
The other  parameters are as follows: $\gamma=1~\mu$eV,
$\gamma'=4\mu$eV, and $\kappa_c=50~\mu$eV ($\kappa_t=62.5~\mu$eV).
Of note, the polariton doublet in (a) is not due to strong coupling, since we are in the weak coupling regime for these simulations (see text). 
}
\label{fig:2}
\vspace{0.25cm}
\includegraphics[width=0.95\columnwidth]{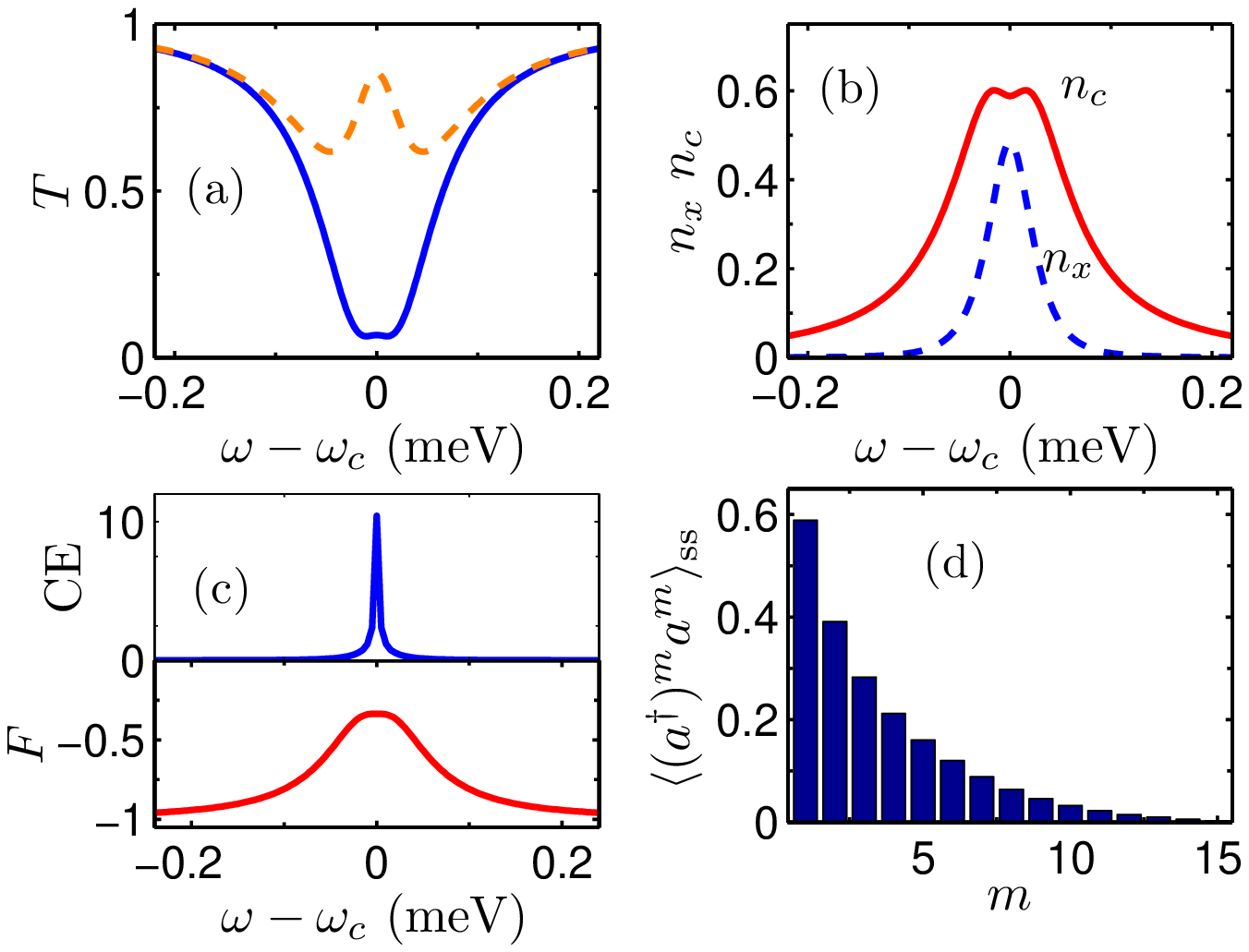}
\vspace{-0.15cm}
\caption{(Color online)
As in Fig.~\ref{fig:2}, but with the
larger pump field $\eta_c=2.5\,g$.
We clearly see that multiphoton correlations
are needed, even for this weakly coupled system ($g\ll\kappa_t$).
}
\label{fig:3}
\end{figure}

We first investigate the system with no acoustic phonon coupling---apart from the ZPL broadening (i.e., no ${\cal L}_{\rm ph}$ process and $\braket{B}=1$). For calculations,  we  use  material parameters that correspond closely to those in experiments: $\tilde\gamma_{x}=1~\mu$eV, $\kappa_c=4\kappa_0=50~\mu$eV,
  $ \gamma^{\prime}=4~\mu$eV (unless stated otherwise), and we  will study various values of $g$ and $\eta_c$.

In Fig.~\ref{fig:2} we study the weak-coupling regime, with $g=20~\mu{\rm eV}\approx0.32\,\kappa_t$,
using a fairly weak  excitation field of $\eta_x=0.5\,g$. This field value
was chosen to be small enough that the cavity population
is significantly lower than 0.1, but large enough to see a breakdown
of the WEA. We have also confirmed that this value of $g/\kappa$ yields no vacuum Rabi splitting.
In Fig.~\ref{fig:2}(a) we show the transmission versus detuning
with (dashed) and without the WEA (solid); a few comments are in order:
($i$) we confirm that the polariton doublet appears even though we are {\em not}
in the strong coupling regime (also see  Refs.~\onlinecite{Waks:PRL06,Auffeves:PRA07,Rakher:PRL09}); ($ii$) the WEA breaks down with already
qualitative differences of more than 40\% near $\omega\approx\omega_c$;
($iii$) with the chosen value of $g$, the region of ``transparency'' is notably very weak, which is a consequence of the finite
QD broadening (through $\tilde\gamma$ and $\gamma'$). This latter observation can be contrasted with the results this can be contrasted to the 
results in Ref.~\onlinecite{Auffeves:PRA07} where such broadenings were not included; these ZPL broadenings are essential to include for a realistic QD system. In Fig.~\ref{fig:2}(b), we show the (numerically-exact) 
exciton and cavity-mode populations, confirming that the
largest cavity population is well below 0.1;
however, we note that  the fundamental condition for the WEA is not a low number of photons, but a negligible excitation of the dot---and the exciton population is evidently no longer negligible. In Fig.~\ref{fig:2}(c), we display the  correlation error, CE, 
and the Fano factor, $F$.
Even at these weak drives and small $g$ (weak coupling regime), it is clear
that a semiclassical approximation can fail, especially
near $\omega \approx \omega_c$ where the percentage error
of assuming $\braket{a^\dagger\sigma^-}_{\rm ss} \approx \braket{a^\dagger}_{\rm ss}\braket{\sigma^-}_{\rm ss}$
is as much as 50\%.
In Fig.~\ref{fig:2}(d), we  plot the corresponding cavity-photon moments
$\braket{(a^\dagger)^m a^m}_{{\rm ss}}$ for the first few
photon states, which suggest that only the first two photon number states are excited. 


\begin{figure}[b!]
\centering
\includegraphics[width=0.95\columnwidth]{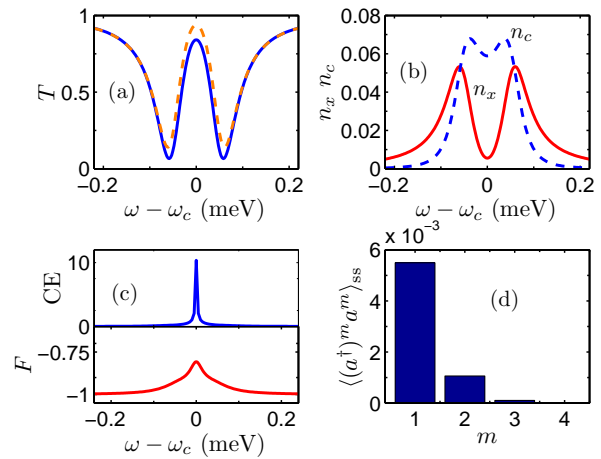}
\vspace{-0.15cm}
\caption{(Color online)
As in Fig.~\ref{fig:2}, but for an intermediate-to-strongly coupled system, with  $g=60~\mu$eV $\approx \,\kappa_t$ and  $\eta_c=0.25\,g$.
The other  parameters are
the same as in Fig.~\ref{fig:2}.
}
\label{fig:4}
\end{figure}

To better probe the nonlinear quantum  aspects of this dot-cavity coupling regime,
in Fig.~\ref{fig:3} we increase the pump value to $\eta_c=2.5\,g$.
Here the WEA breaks down dramatically, as shown by Fig.~\ref{fig:3}(a); Figs.~\ref{fig:3}(c)-(d) further confirm that  we are accessing a regime
where both the WEA and the semiclassical approximations may fail, even for a weakly coupled system. We are thus already in the anharmonic cavity-QED regime.
Figure 3(b) shows the corresponding populations.
\begin{figure}[t!]
\centering
\vspace{0.25cm}
\includegraphics[width=0.95\columnwidth]{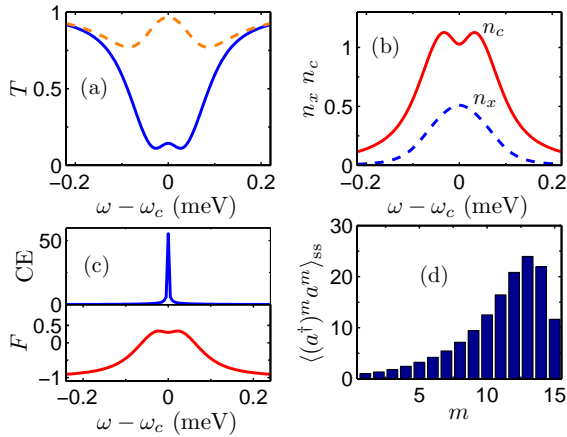}
\vspace{-0.15cm}
\caption{(Color online)
As in Fig.~\ref{fig:4}, but with the
larger pump field $\eta_c=1.2\,g$.
}
\label{fig:5}
\end{figure}


Next, we investigate a QD-cavity system in the intermediate-to-strong coupling regime, with $g=60~\mu{\rm eV}\approx \kappa_t$, and
 $\eta_c=0.25\,g.$
In Fig.~\ref{fig:4}(a), we  show the transmission
with (dashed) and without (solid) the WEA; we also show the  populations [\ref{fig:4}(b)],  the semiclassical error and the Fano factor [\ref{fig:4}(c)].
 For this relatively weak drive,
we again observe noticeable differences in the WEA and non-WEA
predictions; we also recognize that higher-order
photon moments are significant already
in the $m=2$ photon state [Fig.~4(b)].

With increasing drives, 
namely when $\eta_c=1.2\,g$,
Fig.~5  confirms that the 
differences between the WEA and multiphoton calculations
are even more dramatic (as expected), and the on-resonance transmission becomes much smaller
with the nonlinear drive~\cite{Auffeves:PRA07}. Indeed, we recognize that
multiphoton effects are now fairly profound, easily exciting the 
first 15 photon states. This means that we are now
accessing the first 30 states of the JC ladder, with  modest exciton-cavity coupling rates (i.e., $g \sim \kappa_t$). 

\subsection{Influence of acoustic phonon coupling}

For the phonon bath calculations,  we  use  material parameters
for InAs QDs~\cite{roy_hughes,hughes1}, with  $\omega_b=1~$meV
  and $\alpha_p/(2\pi)^2=0.06\,{\rm ps}^2$.
The phonon ME model considers a bath at a temperature of $T=20~$K, resulting in $\braket{B}(20\,{\rm K})=0.73$.
We also now consider a QD-cavity system in the strong coupling regime, with $g=120~\mu{\rm eV}\approx 2\,\kappa_t$, using
a low cavity pump rate of $\eta_c=0.15\,g.$

\begin{figure}[b!]
\centering
\includegraphics[width=0.95\columnwidth]{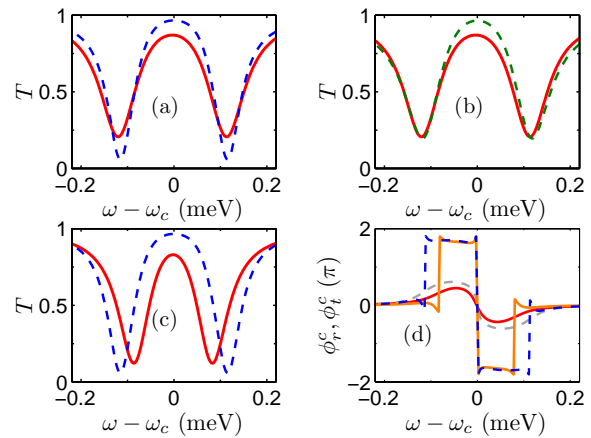}
\vspace{-0.15cm}
\caption{(Color online)
Transmission/reflection characteristics of a  weakly excited
strongly-coupled coupled system, $\eta_c=0.15\,g$, with $g=120~\mu$eV $\approx 2\,\kappa_t$. The peak exciton and cavity photon populations are both below
0.1 (not shown). For these calculations we
also consider the effect of acoustic phonon scattering
at a bath temperature of $T=20~$K.
(a) Transmission with (solid) and without (dashed) 
incoherent photon scattering terms, included through ${\cal L}_{\rm ph}$; here we ignore coherent renormalization effects, i.e.,
$\braket{B}=1$.
(b) Transmission as in (a), but now teh non-phonon case has  $\gamma'\rightarrow 1.6\,\gamma'$
to try and mimic the influence of incoherent phonon scattering.
(c) Transmission with (solid) and without (dashed) 
incoherent scattering and a coherent phonon reduction in $g\rightarrow\braket{B}g$ with $\braket{B}=0.73$. (d) Conditional phase gate, where the larger phase changes are $\phi^c_t$  and the smaller phase change is $\phi^c_r$; solid and dashed lines show results with and without phonon coupling, respectfully. All other system and material parameters are
the same as in Fig.~\ref{fig:2}, and no WEA is made (the semiclassical  approximation breaks down dramatically in this strong coupling regime).             
}
\label{fig:6}
\end{figure}

From the theory described in Sec.~\ref{theory}, electron-phonon scattering is seen to manifest in a coherent renormalization in $g\rightarrow \braket{B}g$ as well as mediate incoherent scattering between the exciton and cavity. To better highlight the
role of incoherent scattering separately, we first set $\braket{B}=1$
(i.e., we neglect the coherent phonon effects) for the calculations  in Fig.~\ref{fig:6}(a) and Fig.~\ref{fig:6}(b).
Figure~\ref{fig:6}(a) shows the transmission results with (solid) and without (dashed) incoherent phonon scattering. We recognize a clear change in the maximum and minimum transmission regions and a qualitative reshaping of the spectral profile. It is common to try and partially mimic the effects of phonon scattering by using an effective $\gamma'$ to fit the data.
To highlight the differences with such a simple Lorentzian coupling approach, in Fig.~\ref{fig:6}(b) we try to fit the phonon model by increasing $\gamma'\rightarrow 1.6\gamma'$, which
naturally broadens the ZPL through an increase of the  (Lorentzian model) pure dephasing process. However, comparing the
dashed and solid curves of Fig.~\ref{fig:6}(b), we see that the features near the center peak and the edges of the graph are noticeably different; in particular, a broadened  ZPL  incorrectly widens the Lorentzian tails and misses the reduction of the center transmission peak. In addition, a change in pure dephasing does not obtain
the coherent reduction of $g$ ($g\rightarrow 0.73g$ since $\braket{B}=0.73$ at 20\,K), which is an important temperature-dependent effect
that we include in Fig.~\ref{fig:6}(c); this marked reduction in the Rabi
splitting will increase with temperature, which is important to note especially if trying to fit experimental data that is probed via temperature tuning. For lower bath temperatures, the
on-resonance case can also be asymmetric~\cite{Frank:PRB08}.
We further note that because we are exciting through the cavity mode,
the effects of the drive on excitation-induced dephasing are significantly suppressed, which is in this  contrast to driving through the
QD exciton~\cite{stuttgart_prl,roy_hughes,nazir2,ramsay2}.

Finally, we study a {conditional} phase gate. We define the conditional phase
 through $\phi^c_{t/r}=\phi^d_{t/r}-\phi^0_{t/r}$, where
$\phi^d_{t/r}$ includes the dot resonance and
$\phi^0_{t/r}$ is the phase without the dot.
Using a semiconductor micropillar system, conditional phase shifts of around 0.03~rad
were recently observed~\cite{Young:arXiv10}.
 Figure~\ref{fig:6}(d) shows the conditional phases, with (solid) and without (dashed) phonon
 coupling.
 Near the spectral regions near $\pm 0.1~$meV, about $4\pi$ transmission phase change is possible
{\em conditioned} upon the dot exciton being at the frequency regime---which can be tuned
for example by applying a field-induced Stark shift. In reflection, conditional phase changes of
around $\pm \pi/4$ are found to be possible; note that the one photon results (i.e., the WEA) tend to overestimate this value.
Including the coupling to the  phonon bath is seen to qualitatively change the phase characteristics, and
we
stress that there is no
 effective
 $\gamma'$ which could mimic the same phase trends with phonon bath coupling;
  the large negative phase of transmission actually widens with increasing
$\gamma'$, instead of narrowing then eventually  disappears for increasing drives.
The phase gate characteristics can be optimized further by changing the phonon bath temperature and by changing the exciton-cavity detuning.

\section{Conclusions}
\label{conclusions}

We have presented a semiconductor ME formalism that can accurately
simulate coherent input/output coupling of open-system semiconductor
cavity-QED
systems
such as planar photonic crystals and micropillar cavities. We investigated the role of quantized multiphoton effects
and pointed out the possible failure of the WEA,
which is shown to  fail even for low
input powers and small mean cavity photon numbers (much lower than 0.1). 
For increasing field strengths, the possible failure of the semiclassical approach is also highlighted. We have further shown that
coupling to an acoustic phonon bath causes considerable qualitative changes to the
light propagation characteristics than is modeled by a simple pure dephasing process. Finally, we  used this model to  simulate a conditional phase gate.

\section*{Acknowledgments}
This work was supported by the National Sciences and
Engineering Research Council of Canada. We thank
H. J. Carmichael
for useful discussions and
acknowledge
 use of the quantum optics
toolbox~\cite{QOToolbox}.

\end{document}